\documentclass[aps,prl,letterpaper,11pt,twoside,tightenlines,nofootinbib,showpacs,preprint,onecolumn]{revtex4}
\usepackage{graphicx}
\usepackage[sort&compress]{natbib}
\usepackage{latexsym}
\usepackage{epsfig}
\begin{document}
\newcommand{\eg}{{\it e.g.}}
\newcommand{\etal}{{\it et. al.}}
\newcommand{\ie}{{\it i.e.}}
\newcommand{\be}{\begin{equation}}
\newcommand{\dd}{\displaystyle}
\newcommand{\ee}{\end{equation}}
\newcommand{\bea}{\begin{eqnarray}}
\newcommand{\eea}{\end{eqnarray}}
\newcommand{\bef}{\begin{figure}}
\newcommand{\eef}{\end{figure}}
\newcommand{\bce}{\begin{center}}
\newcommand{\ece}{\end{center}}
\def\lsim{\mathrel{\rlap{\lower4pt\hbox{\hskip1pt$\sim$}}
    \raise1pt\hbox{$<$}}}         
\def\gsim{\mathrel{\rlap{\lower4pt\hbox{\hskip1pt$\sim$}}
    \raise1pt\hbox{$>$}}}         

\title{Thermal Hadronization, Hawking-Unruh Radiation and Event Horizon in QCD}

\author{P.CASTORINA}

\address{Physics Department, Catania University,\\
Via Santa Sofia 64 Catania , Italy\\
$^*$E-mail: paolo.castorina@ct.infn.it}

\begin{abstract}
Because of colour confinement, the physical vacuum forms an event 
horizon for quarks and gluons; this can be crossed only by quantum 
tunneling, i.e., through the QCD counterpart of Hawking radiation by 
black holes. Since such radiation cannot transmit information to the 
outside, it must be thermal, of a temperature determined by the strong 
force at the confinement surface, and it must maintain colour neutrality. 
The resulting process provides a common mechanism for thermal hadron 
production in high energy interactions, from $e^+e^-$ annihilation 
to heavy ion collisions. The analogy with black-hole event horizon suggests
a dependence of the hadronization temperature on the baryon density.   
\end{abstract}

\pacs{25.75.-q}
\maketitle

\section{Introduction}\label{aba:sec1}

Over the years, hadron production studies in a variety of high energy 
collision experiments have shown a remarkably universal feature. From 
$e^+e^-$ annihilation to $p-p$ and $p-\bar p$ interactions and further 
to collisions of heavy nuclei, covering an energy range from a few GeV up 
to the TeV range, the production pattern always shows striking thermal 
aspects, connected to an apparently quite universal temperature around
$T_H \simeq 160 - 190$ MeV \cite{Hagedorn,species}. 

What is the origin of this thermal behaviour? While high energy heavy ion 
collisions involve large numbers of incident partons and thus could allow 
invoking some ``thermalisation'' scheme through rescattering, in $e^+e^-$ 
annihilation the predominant initial state is one energetic $q \bar q$ pair, 
and the number of hadronic secondaries per unit rapidity is too small to 
consider statistical averages.

A further piece in this puzzle is the observation that the value of the
temperature determined in the mentioned collision studies is quite similar 
to the confinement/deconfinement transition temperature found in lattice 
studies of strong interaction thermodynamics \cite{T-c}. While hadronization 
in high energy collisions deals with a dynamical situation, the energy loss 
of fast colour charges ``traversing'' the physical vacuum, lattice QCD 
addresses the equilibrium thermodynamics of unbound vs.\ bound colour
charges. Why should the resulting critical temperatures be similar or even
identical?  

In ref.(4), which is summarized in this contribution 
( see also ref.(5)),
 these hadronization phenomena are considered as  the QCD counterpart
of the Hawking radiation emitted by black holes (BH) \cite{Hawking}.
BHs provide a gravitational form of confinement that
was quite soon  compared to that of colour confinement
in QCD \cite{castor,Salam},  where coloured constituents are confined to 
``white holes'' (colourless from the outside, but coloured inside).

The main results in  ref.(4) are:

\begin{itemize}
\item{Colour confinement and the instability of the physical vacuum under 
pair production form an event horizon for quarks, allowing a transition 
only through quantum tunnelling; this leads to thermal radiation of a 
temperature $T_H$ determined by the string tension.}
\item{Hadron production in high energy collisions occurs through a
succession of such tunnelling processes. The resulting cascade is a
realization of the same partition process which leads to a limiting
temperature in the statistical bootstrap and dual resonance models.}
\item{In kinetic thermalization, the initial state information is
successively lost through collisions, converging to a time-independent  
equilibrium state. In contrast, the stochastic QCD Hawking radiation 
is ``born in equilibrium'', since quantum tunnelling {\sl a priori} 
does not allow information transfer.} 
\item{The temperature $T_H$ of QCD Hawking radiation depends only on 
the baryon number and the angular momentum of the deconfined system.
The former provides the dependence of $T_H$ on the baryochemical 
potential $\mu$, while the angular momentum pattern of the radiation
allows a centrality-dependence of $T_H$ and elliptic flow. In particular the $\mu$ dependence of $T_H$, 
will be discussed  in more details in sec. 4} 

\end{itemize}

\section{Thermal production pattern}

Let us first summarize the thermal production pattern in elementary collisions,
$e^+e^-, p p,  \bar p p,..$ and in nucleus-nucleus scattering.

The partition function of an ideal resonance gas is given by
{\begin{equation}
\ln Z(T) = V \sum_i {d_i \over (2\pi)^3}~\! \phi(m_i,T)
\end{equation}}
\hskip-0.3cm
where $d_i$ is the degeneracy factor and $\phi(m_i,T)$ is the Boltzmann factor
\begin{equation}
\phi(m_i,T)= 4\pi m_i^2 T K_2(m_i/T)
\end{equation} 
Therefore the relative abundances of the species $i$ and $j$ turns out 
{\begin{equation}
{N_i\over N_j}= {d_i \phi(m_i,T) \over d_j \phi(m_j,T)}
\end{equation}}
and for transverse energy larger than $T$  
{\begin{equation}
{dN \over dp_T^2} \sim \exp{-{1\over T}~\!\sqrt{m_i^2 + p_T^2}}
\end{equation}}

In elementary collisions the statistical hadronization model \cite{shm} 
fits the data on the species abundances by two parameters: $T$ and $\gamma_s$,
that describes the strangness suppression.
 
For LEP data at $\sqrt{s}= 91.2$ Gev \cite{species}, $T=170 \pm 3 \pm 6$ Mev and $\gamma_s= 0.691 \pm 0.053$
where the systematic error is obtained by varying the resonansce gas scheme and the contributing resonances.
The PEP-PETRA data at differente energies , $ 14 <\sqrt{s} < 45$ can be fitted \cite{species}
with an average temperature
 $T = 165 \pm 6$ Mev and $\gamma_s \simeq 0.7 \pm 0.05$. The $p p$ SPS data at energies   $ \sqrt{s}= 19, 23.8 , 26$ Gev
give $T=162.4 \pm 1.6$ Mev and  $\gamma_s \simeq 0. \pm 0.036$ \cite{species}. The other data for $K^+ p$ and $\pi^+ p$ scattering at energies 
close to SPS one  and for $ \bar p p$ at larger energy can be fitted by similar values.

The fitted values of the temperature are depicted in fig. 1 \cite{cisco}.

\begin{figure}[h]
{\psfig{file=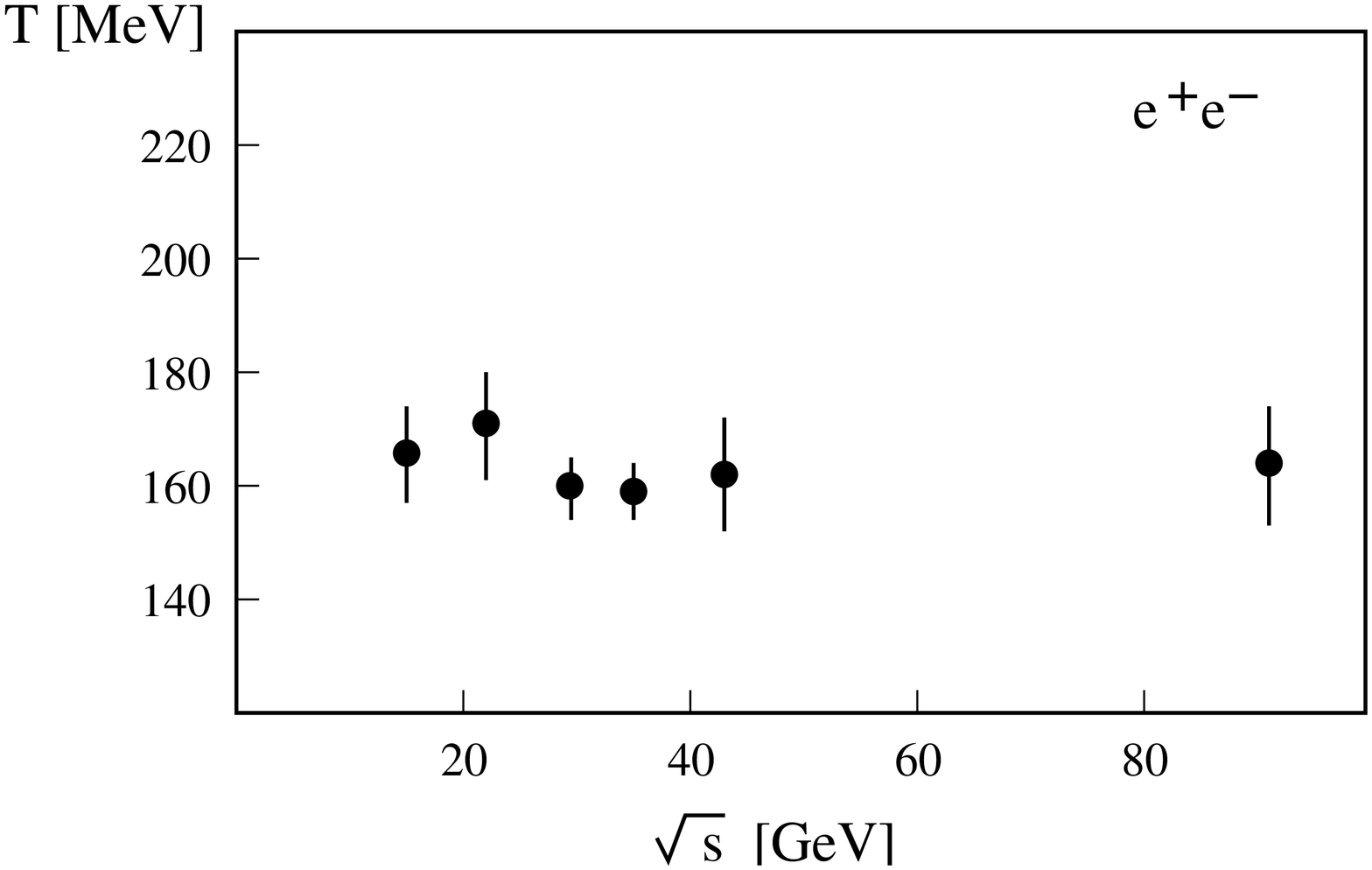,width=9cm}}
\hskip1cm{\psfig{file=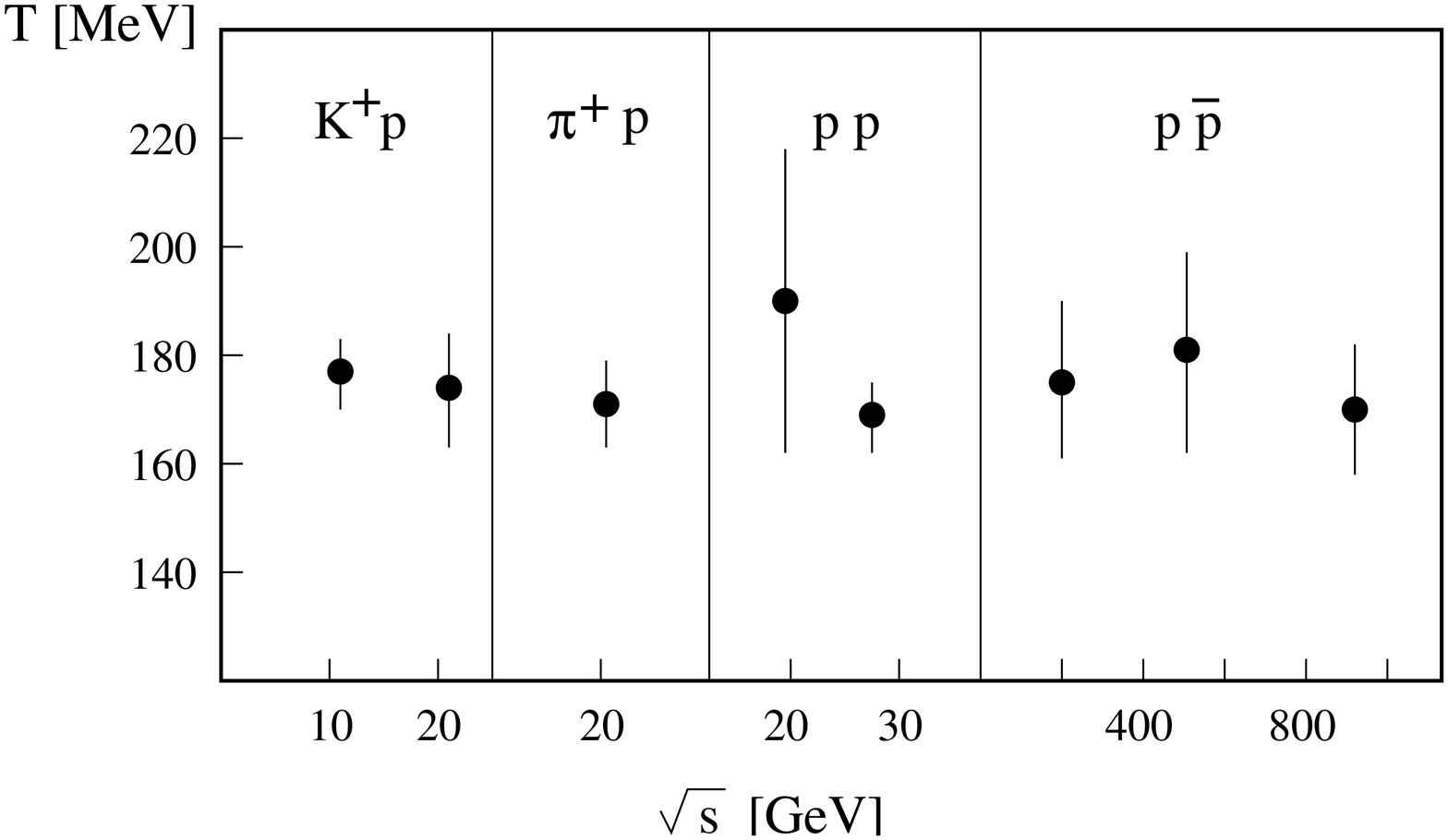,width=10cm}}
\caption{Hadronization temperature at large energy for various elementary collisions}
\end{figure}

Therefore there is an universal hadronization temperatute $T_H = 170 \pm 10-20$ Mev which
is independent on the species, on $\sqrt{s}$ and on the incident configuration. Moreover,    
also the transverse momentum spectra in elementary collisions can be fitted by the same value $T_H$ \cite{species}.

In heavy ion collisions there is a new parameter which describes the finite baryon density, i.e. 
the baryon chemical potential $\mu_B$. 
The fits of the species abundances at high energy ( peak SPS and RHIC) give  \cite{species} :

- {$T_H= 168 \pm 2.4 \pm 10$ Mev; $\mu_b= 266 \pm 5 \pm 30$ Mev at $\sqrt{s}= 17$ Gev for (SPS) Pb-Pb},

- {$T_H =168 \pm 7 $ Mev;  $\mu_b= 38 \pm 11 \pm 5$ Mev at $\sqrt{s}= 130$ Gev for (RICH) Au-Au at y=0},

- {$T_H =161 \pm 2 $ Mev;  $\mu_b= 20 \pm 4 $ Mev at $\sqrt{s}= 200$ Gev for (RICH) Au-Au}.

In conclusion, hadron abundances in all high energy collisions ( $e^+e^-$ annihilation,hadron-hadron and heavy ion
collisions) are those of an ideal resonance gas at universal temperature
$T_H \simeq 170 \pm 10-20 \phantom{.}$ Mev.

\section{Event horizon and Hadronization}

The idea that a thermal medium, with a kinetic thermalization by
multiple  partonic interaction, has been produced  in the collisions could explain the previous phenomenon 
for a nucleus-nucleus scattering but does not work for  $e^+e^-$ and hadron-hadron scattering. 

One has to look for an universal,
``non-kinetic'' thermalization mechanism. 

Indeed, in gravitation there is a  well known example:
the BH Hawking radiation has a  thermal radiation spectrum due to tunnelling through the event horizon and the Hawking temperature
is given by $T_{BH}= 1/8 \pi G  M$, where $M$ is the (Schwarzschild) BH mass and $G$ is the Newton constant  \cite{Hawking}.

Therefore the conjecture \cite{cks} is that colour confinement and hadronization mechanism are in strong analogy with
BH physics and event horizon. 

There are many reasons to believe that color confinement can be described by a color horizon 
in QCD because the theory is non linear and therefore it has  an effective curved geometry \cite{Novello,dima} 
However, since we discuss the hadronization mechanism, is better to consider the Unruh effect. 

As shown by Unruh \cite{Unruh}, systems with uniform acceleration,$a$, have an event horizon and see a thermal bath
with temperature $T_U = a/2 \pi$. For a particle of mass,$m$, in uniform acceleration
the equation of motion is solved by the parametric form  
\begin{equation}
x = {1\over a} \cosh a\tau ~~~~~ t= {1\over a} \sinh a \tau,
\label{rindler}
\end{equation}
where $a=F/m$ denotes the acceleration in the instantaneous rest frame of 
$m$, and $\tau$ the proper time, with $d\tau=\sqrt{1-v^2} dt$. The resulting
world line is shown in fig.2 with the 
event horizon beyond which $m$ classically 
cannot pass. The only signal the observer can detect as consequence of the
passage of $m$ is thermal quantum radiation of temperature 
\begin{equation}
T_U = {a \over 2 \pi}.
\label{T-U}
\end{equation}
In the case of gravity, $a$ is the ``surface gravity''(i.e. the acceleration at the  horizon),
 $a=1/(4~G~\! M)$,  and hence one recovers the Hawking temperature.
\begin{figure}[h]
\centerline{\psfig{file=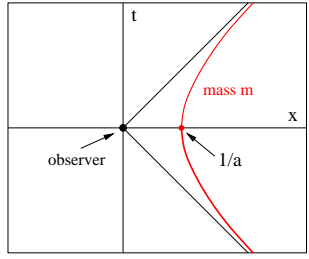,width=4cm}} 
\caption{Hyperbolic motion}
\label{W-L}
\end{figure}

In summary, the acceleration leads to a classical turning point and
hence to an event horizon, which can be surpassed only by quantum tunnelling 
and at the expense of complete information loss, leading to thermal radiation 
as the only allowed signal.

On the other hand,at quantum level,it is well known that in a strong field
the vacuum is unstable against pair production \cite{Schwinger}. For example, in
$e^+e^-$ annihilation a $\bar q q$ pair is initially produced and when the  linear potential
is such that  $\sigma x > \sigma x_Q \equiv 2m$
the string connecting  $\bar q q $ breaks and the color neutralization induces an effective
quantum event horizon (see figs 3,4)

\begin{figure}[h]
\centerline{\psfig{file=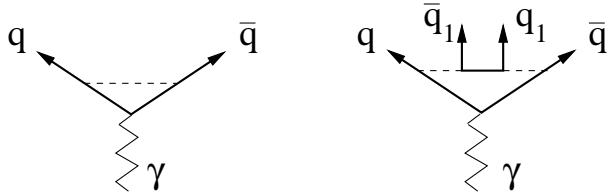,width=8cm}}
\caption{Pair production in $e^+e^-$}
\end{figure}

\begin{figure}[h]
\centerline{\psfig{file=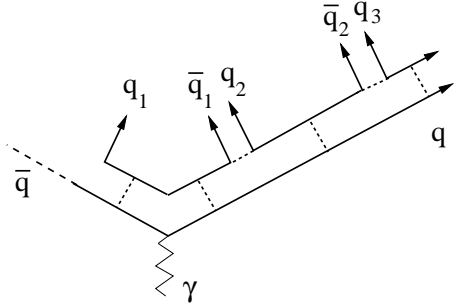,width=6cm}}
\caption{Hadronization  as a tunnelling process}
\end{figure}

The  $\bar q q $ flux tube has a thickness given by \cite{cks} 
\begin{equation}
 r_T \simeq \sqrt{2 \over \pi \sigma}
\end{equation}
and the  $\bar q_1  q_1$ is produced at rest in $e^+e^-$ cms but with a transverse momentum
\begin{equation}
k_T \simeq {1 \over r_T} \simeq  \sqrt{\pi \sigma \over 2}.
\end{equation}

The acceleration (or deceleration) associated with the  string breaking and color neutralization
mechanism turns out to be \cite{cks}
\begin{equation}
a  \simeq 2~\!k_T \simeq \sqrt{2 \pi \sigma}
\simeq 1.1~{\rm GeV},
\end{equation}
which leads to
\begin{equation}
T_q = {a\over 2 \pi} \simeq \sqrt{\sigma \over 2 \pi} \simeq 180~{\rm MeV}
\label{T-H}
\end{equation}
for the hadronic Unruh temperature. It governs the momentum distribution
and the relative species abundances of the emitted hadrons.

Notice that the previous hadronization mechanism can be described by
saying that {$\bar q_1$} reaches the {$q_1 \bar q_1$} event horizon and 
tunnels to become {$\bar q_2$}. The  
emission of hadron {$\bar q_1 q_2$} can be considered  as Hawking radiation.

\section{ Vacuum Pressure and Baryon Density}

\vskip0.5cm

It is interesting  to consider the extension of the previous mechanism
in the case of systems with a net baryon number,i.e with a new ``charge''
which can modify the tunnelling process and the Hawking-Unruh hadronization temperature.

In the BH case the effect of a total charge $Q$ changes the Hawking temperature according to the
formula  ( see for example \cite{Ruf})   
\begin{equation}
T_{BH}(M,Q) = T_{BH}(M,0)~\!\left\{{4~\sqrt{1 - Q^2/ GM^2} 
\over (1 + \sqrt{1- Q^2/GM^2}~\!)^{~\!2}}\right\}; 
\end{equation}
Note that with
increasing charge, the Coulomb repulsion weakens the gravitational field
at the event horizon and hence decreases the temperature of the corresponding
quantum excitations. As $Q^2 \to GM^2$, the gravitational force is fully
compensated. The crucial quantity here is the
ratio $Q^2/GM^2$ of the overall Coulomb energy, $Q^2/R$, to the overall 
gravitational energy, $GM^2/R$. Equivalently, $Q^2/GM^2 = P_Q/P_G$
measures the ratio of inward gravitational pressure $P_G$ at the event 
horizon to the repulsive outward Coulomb pressure $P_Q$.  

In QCD, we have a ``white'' hole containing coloured quarks, confined 
by chromodynamic forces or, equivalently, by the pressure $B$ of the
physical vacuum. If the system has a non-vanishing overall baryon number,
there will be a Fermi repulsion between the corresponding quarks, and this 
repulsion will provide a pressure $P(\mu)$ acting against $B$, with 
$\mu$ denoting the corresponding quark baryochemical potential. We thus
expect a similar reduction of the hadronization temperature as function
of $\mu$. To quantify this aspect let us consider, at $T=0$,
the Fermi pressure,
$P=d_g \mu^4/24 \pi^2$, where 
$d_g$ is the degeneracy factor,  versus the QCD vacuum pressure 
due to the gluon condensate $B=<(\alpha_s / \pi) G_{\mu \nu}^2>$
which is a decreasing ( largely unknown) function of the  baryon density. By following the 
analysis of ref. \cite{bcz} for the
dependence  of the gluon condensate on the baryon number,
the critical  density, where $B$ balances the Fermi pressure, turns out 
about $n_c = 5-6 n_0$, with 
$n_0$  the nuclear saturation density.

A slightly different result is obtained in the description of  deconfinement by
percolation \cite{percolation} : $n_c \simeq 4 n_0$.

In the intermediate
region, where both $T$ and $\mu$ are finite, we want to compare the effect
of the Fermi repulsion to the vacuum pressure through the Hawking-Unruh
form, i.e., we replace $Q^2/GM^2$ in eq.(11) by $(\mu/\mu_0)^4$, giving 
\begin{equation}
T(\mu)/T_0 = {4 \sqrt{1-(\mu/\mu_0)^4} \over (1+\sqrt{1-(\mu/\mu_0)^4})^2}.
\end{equation}
The resulting behaviour of $T(\mu)$ is shown in Fig.5  
for $\mu_0$ corresponding to the previous $n_c \simeq 4 - 6$ $ n_0$. 
In the same figure is shown the results obtained by using
$(\mu/\mu_0)^2$ rather than $(\mu/\mu_0)^4$ in eq.(12).
 In both cases the function
remains rather flat up to large value of $\mu$.

\begin{figure}[h]
\centerline{\psfig{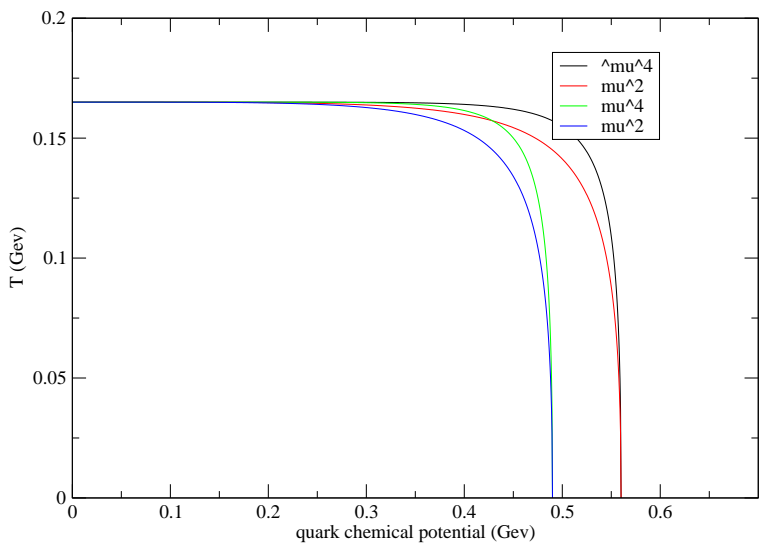}}
\caption{T dependence on $\mu$ by eq.(12) with  $(\mu/\mu_0)^n$
 for $n=2,4$}
\end{figure}

Clearly this approach is overly simplistic, since it reduces the effect
of the additional quarks to only their Fermi repulsion. A more general way 
of addressing the problem would be to introduce an effective $\mu$-dependence 
of the string tension.


\end{document}